\documentclass{iopart}
\pdfoutput=1

\usepackage{iopams}

\usepackage{color}
\usepackage{dcolumn}
\usepackage{graphicx}
\usepackage[utf8]{inputenc}
\usepackage{latexsym}
\usepackage{times}

\newcommand{\scri}{\ensuremath{\mathcal{J}^+}}

\newcommand{\ie}{\textit{i.e.,\ }}
\newcommand{\eg}{\textit{e.g.,\ }}

\renewcommand{\th}{\tilde{h}}
\newcommand{\news}{\ensuremath{\mathcal{N}}}
\newcommand{\sinc}{\ensuremath{\rm sinc}}

\definecolor{rgb_blue}{rgb}{0.0,0.0,0.6}
\definecolor{rgb_green}{rgb}{0.0, 0.6, 0.0}
\definecolor{rgb_red}{rgb}{0.7,0.,0.}

\newcommand{\eqref}[1]{Eq.~(\ref{#1})}

\begin{document}

\title{Notes on the integration of numerical relativity waveforms}

\author{Christian Reisswig}
\address{
  Theoretical Astrophysics Including Relativity,
  California Institute of Technology,
  Pasadena, CA 91125, USA
}

\author{Denis Pollney}
\address{
  Departament de F\'isica,
  Universitat de les Illes Balears,
  Palma de Mallorca,
  E-07122, Spain
}

\date{\today}

\begin{abstract}
  A primary goal of numerical relativity is to provide estimates of
  the wave strain, $h$, from strong gravitational wave sources, to be
  used in detector templates.  The simulations, however, typically
  measure waves in terms of the Weyl curvature component, $\psi_4$. 
  Assuming Bondi gauge, transforming to the strain $h$
  reduces to integration of $\psi_4$ twice in time. 
  Integrations performed in either the time or frequency domain, however, lead to secular non-linear drifts in
  the resulting strain $h$. These non-linear drifts are not explained by the two unknown integration
  constants which can at most result in linear drifts.
  We identify a number of fundamental difficulties which can arise from integrating
  finite length, discretely sampled and noisy data streams. 
  These issues are an artifact of post-processing data. They are
  independent of the characteristics of the original simulation, such as
  gauge or numerical method used.
  We suggest, however, a simple procedure for
  integrating numerical waveforms in the frequency domain, which is
  effective at strongly reducing spurious secular non-linear drifts in the
  resulting strain.  
\end{abstract}


\maketitle

\section{Introduction}

With the advent of gravitational wave detector experiments, the
concept of metric strain has been elevated from a theoretical result
of the linearized Einstein equations, to a genuine physical observable
which will be directly measured for the first time in the coming
years.  Since they are weak, foreknowledge of the expected signals
will greatly aid the initial detection and subsequent understanding of
measurements.  Thus, a number of major efforts are going into constructing
high-precision models of dynamical spacetimes, in order to determine
what the detectors will see from strongly radiating burst sources such
as binary black hole and neutron star mergers.

Numerical models have achieved some timely successes
in recent years~\cite{Pretorius:2005gq, Campanelli:2005dd,
  Baker:2005vv}, so that the final orbits, merger and ringdown of
binary black holes can be modeled with high numerical
accuracy. Templates for these waveforms are being constructed by
matching the numerical results to post-Newtonian inspirals, for
instance using the effective one-body approach~\cite{Damour:2007vq,
  Damour:2008te, Damour:2009kr, Buonanno:2009qa, Pan:2009wj}, or
purely phenomenological models~\cite{Ajith:2007qp, Ajith:2007kx,
  Ajith:2009bn, Santamaria:2010yb}. Meanwhile, these models are
already being tested in detector search pipelines~\cite{Aylott:2009ya, Aylott:2009tn,
  ninjaweb}.

In constructing templates, the natural observable is the one which is
also measured by the detectors, namely the gravitational wave strain $h\equiv h_+-ih_\times$,
decomposed into `$+$' and `$\times$' polarizations in the
transverse-traceless gauge. However, this is typically
not the quantity which is directly computed in numerical simulations,
where the output of numerical simulations is more
usually in the form of curvature tensor components, or
Zerilli-Moncrief-type variables defined relative to a background.
The results are then
transformed to determine the standard $h_+$ and $h_\times$ strain
modes in order to connect to the detector measurements.

Transforming the measured variables to the strain involves some numerical
subtleties. In particular, it has long been
noted that producing a strain, $h$, from the Newman-Penrose curvature
component, $\psi_4$, typically results in a waveform with an
unphysical secular \emph{non-linear} drift (e.g.~\cite{Baker:2002qf, Berti:2007snb}).
The fact that this drift is non-linear indicates that it is not
simply a result of the two unknown integration constants involved in
the transformation. A potential source of the problem may come from the fact
that $\psi_4$ is typically extracted at a finite distance from the gravitating source (\cite{Hannam:2009hh} and references therein).
The strain $h$, however, is related to $\psi_4$ only at an \textit{infinite} distance from the
source hence introducing a systematic ``finite-radius'' error.
Furthermore, the relation between $h$ and $\psi_4$ is strictly only valid in a particular gauge.
This gauge, however, is typically not imposed during the simulations but is given by the
the gauge driver controlling the gauge during the evolution (\cite{Hannam:2009hh} and references therein), and may thus also lead to
secular effects like the observed non-linear drift in the strain.
By eliminating these systematic errors, one would therefore hope to greatly reduce the secular behavior.
Unfortunately, even with the recent possibility of
extracting truly gauge-invariant waveforms at future null
infinity~\cite{Reisswig:2009us, Reisswig:2009rx}, we still observe
non-linear contributions to the drifts on time integration of the results, even though the
measurement is free of gauge and finite-radius errors.  This suggests that
the source of the problem must have different roots.

This paper argues that an important source of unphysical
non-linear drift in numerical computations of gravitational wave
strain lies in the transformation of the measured data (commonly the
Newman-Penrose variable $\psi_4$) to the observable strain $h$, which
generically involves an integration in time. The output of the
numerical simulation is a discretely sampled time series of finite
duration, incorporating some component of unresolved frequencies due
to numerical error. The latter aspect can lead to an uncontrollable
non-linear drift if the integration is performed in the time
domain. An alternative is to perform the integration in the frequency
domain (\eg~\cite{Campanelli:2008nk, Santamaria:2010yb}).
In this case, however, the finite duration of the numerical
signal becomes an issue, as artificial low-frequency components of the
infinite spectrum of a localized-in-time function dominate the
integral. An appropriately chosen band-pass filter improves the
situation greatly. Unfortunately, this can require some complicated
adjustment of parameters, which is difficult to systematize.

We first outline some numerical problems inherent in determining the
strain from gauge-invariant quantities typically used in spacetime
simulations, discussing aspects of time and frequency domain
integrations, and the use of band filters. Finally we arrive at a
new and constructive procedure for performing the required numerical
integrations, \emph{fixed frequency integration} (FFI), which
involves a single parameter related to the lowest physical frequency component of
the wave. The method is effective at reducing secular non-linear drifts in $h$,
while maintaining the energy of the wave. We demonstrate our results
with numerically generated gravitational waveforms, as well as with some
artificial analytic functions in order to gauge the potential errors.


\section{Evaluating gravitational strain from numerical data}

Gravitational waves are dynamic solutions of the nonlinear Einstein
equations, which are most readily described by perturbations of
a fixed background metric:
\begin{equation}
  g_{\alpha\beta} = g^0_{\alpha\beta} + h_{\alpha\beta},
\end{equation}
where $g^0_{\alpha\beta}$ is a fixed background, and $h_{\alpha\beta}$
is a perturbation containing the wave. The
observables measured by a gravitational wave detector are
the strain components, $h_+$ and $h_\times$, in the
transverse-traceless (TT) gauge.  A number of techniques are available
for computing these variables, though usually dependent on some underlying
assumptions regarding the spacetime within a simulation, and
coordinates at some finite distance from the source.

One practical method for evaluating gravitational waves is based on
the extensive work that has been done defining perturbative variables
that are gauge invariant relative to a fixed spherical or axisymmetric
background. Early perturbative studies of black hole spacetimes
~\cite{Regge57, Zerilli70a, Moncrief74} formalized a 1st-order gauge
invariant representation of the variables. These methods have been
applied to numerical relativity simulations for some
years~\cite{Abrahams90, Anninos94a, Abrahams95b}
(see~\cite{Pollney:2007ss} for recent comparisons with $\psi_4$-based
measurement).  Briefly, the formalism defines a set of 1st-order even
and odd-parity gauge invariant variables, $Q^+_{\ell m}$,
$Q^\times_{\ell m}$ describing the metric perturbation.  The $h_+$ and
$h_\times$ components of the strain in the TT-gauge are determined
via:
\begin{equation}
  \label{eq:wave_gi}
  h_+-ih_{\times} =
    \frac{1}{\sqrt{2}r}\sum_{\ell=2}^{\infty}\sum_{m=-\ell}^{\ell}
    \Biggl( Q_{\ell m}^+ -i\int_{-\infty}^t Q^\times_{\ell
    m}(t')dt' \Biggr)\,_{-2}Y^{\ell m}\ ,
\end{equation}
where $r$ is the distance to the source, $t$ the observation
time, and ${}_{-2}Y^{\ell m}$ are spin-2 spherical harmonics.

Alternatively, the curvature can be expressed in terms of Newman-Penrose
(NP) components in a given null frame~\cite{Penrose:1963}. Ideally,
this is performed at null infinity, $\scri$, where the frame and
coordinates can be invariantly specified, and the fall-off of the
curvature is known for asymptotically flat spacetimes.  Procedures for
invariant measures at $\scri$ have recently been developed as a
practical tool~\cite{Reisswig:2009us, Reisswig:2009rx}. However, at a
large but finite distance from the source, accurate measurements can
also be made simply by evaluating the curvature relative to a radially
oriented null frame. The gravitational wave information is determined
either from the asymptotically defined Bondi
news~\cite{Bondi62,Sachs62}~\footnote{See also~\cite{Deadman:2009ds}
  for a recent discussion in the context of 3+1 numerical
  relativity.},
\begin{equation}
  \news = -\Delta \bar\sigma,
\end{equation}
(in the NP notation) or the Weyl curvature component $\psi_4$. Then
the strain is determined by:
\begin{equation} \label{eq:h_from_N}
  h = h_+ - ih_\times 
    = \int^t_{-\infty} dt' \news\,
    = \int^{t}_{-\infty} dt^\prime 
      \int^{t^\prime}_{-\infty} dt^{\prime\prime} \psi_4\,.  
\end{equation}

Significantly, for any of these choices of gauge invariant observable,
$Q^{+,\times}$, $\news$ or $\psi_4$, we only recover the strain, $h$, after one
or multiple integrations in time. One time integral is required in the case of the
perturbative techniques and the Bondi news, $\news$, (as well as the
\emph{strain-rate}, $\dot{h}$, defined in~\cite{Baker:2008}), while
two are required to calculate $h$ from $\psi_4$. In practice,
the integration is not performed from $t=-\infty$, but starts at a particular point, the beginning of the simulation.
This introduces one or two integration constants in \eqref{eq:wave_gi} and \eqref{eq:h_from_N}, respectively.
The integrated result therefore yields at most a \emph{linear} drift, which can easily be removed
by fixing the integration constants, e.g.~by an averaging procedure.

\begin{figure}
  \includegraphics[width=1.\linewidth]{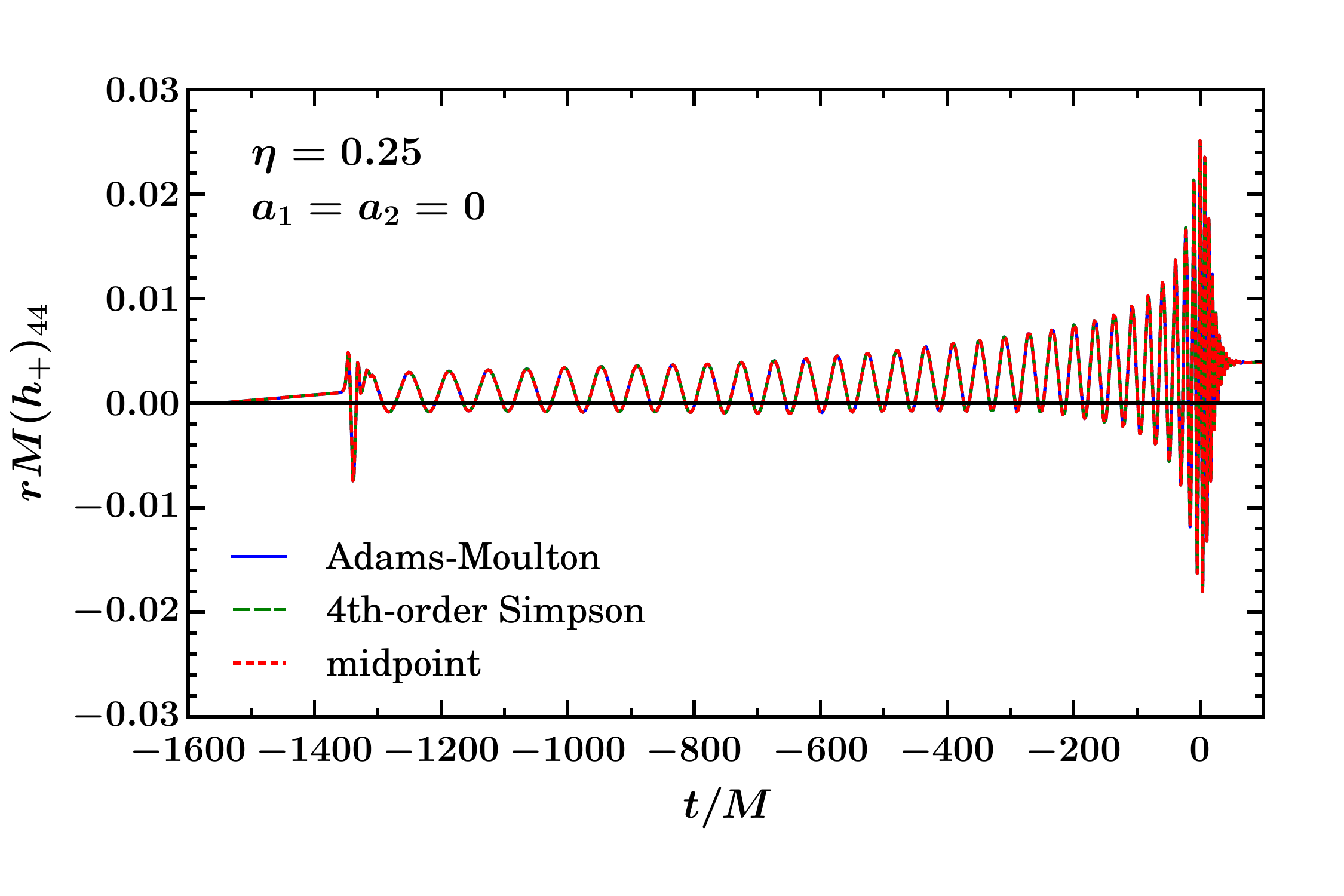}
  \caption{\label{fig:wave_drift} After fixing the integration constants 
    such that linear drifts are removed, a spurious non-linear drift remains in the
    $(\ell,m)=(4,4)$ harmonic mode of $h$ as integrated twice from
    $\psi_4$ of a non-spinning ($a_1=a_2=0$), equal-mass ($\eta=0.25$)
    binary black hole merger simulation.
    For the sake of demonstration, we have chosen a mode
    where the effect is pronounced for this set of numerical data,
    however the dominant $(\ell,m)=(2,2)$ mode shows very similar,
    though more subtle, artifacts. The similar results obtained by an
    Adams-Moulton integration, 4th-order Simpson's rule, and 2nd-order
    midpoint rule, suggest that the drift is independent
    of the integration method. 
  }
\end{figure}

In practice, numerical integration of a time series can be 
performed through standard methods,
for instance, a simple application of Simpson's rule. However, in the
case of gravitational waveform data, after having fixed the integration constants to remove the linear drift,
these procedures tend to introduce a residual \emph{non-linear} drift which is, at best,
a significant nuisance to analysis, but may also be confused with physical modes in which secular drifts are expected \cite{Baker:2002qf, Berti:2007snb}. An example of the problem is shown
in Fig.~\ref{fig:wave_drift}, which plots the $(\ell,m)=(4,4)$
spherical harmonic mode of the strain, $(h_+)_{44}$, determined by
integrating a numerical $(\psi_4)_{44}$ over the last cycles of a non-spinning
 ($a_1=a_2=0$) equal-mass ($\eta=M_1M_2/(M_1+M_2)^2=0.25$) binary black hole merger~\cite{Pollney:2009ut,
  Pollney:2009yz}. We plot the results of time integrations via an
extended 4th-order Simpson's rule, by an Adams-Moulton integration, as
well as a 2nd-order midpoint rule~\cite{Press92}. We have choosen the first integration constant such
that the signal is zero after ringdown. The second integration constant was set to zero.
Whereas we expect
the result to oscillate about zero, in fact we observe a prominent
non-linear drift, which is independent of the numerical integration
method. Similar artifacts have been observed in waveform
computations from different simulations, 
for instance early ringdown results~\cite{Baker:2002qf},
as well as more recent studies~\cite{Berti:2007snb,Boyle:2008ge}. 

One can imagine
a number of systematic sources for the unphysical non-linearities of the drift, resulting from the way
measurements are made within the simulation.  For
measurements at finite radius, the observer location (typically a
sphere at some radius defined by grid coordinates) changes over time
for the dynamical coordinate conditions which are in common use. 
However, the waveforms plotted in
Fig.~\ref{fig:wave_drift} are measured at $\scri$, via characteristic
extraction \cite{Reisswig:2009us, Reisswig:2009rx}. As such, they
should be immune to local coordinate effects, and indeed,
examination of $\psi_4$ using different worldtube data and different resolutions
reveal that the differences in $\psi_4$ converge to zero \cite{Reisswig:2009us, Reisswig:2009rx}.
This suggests that the source of the problem must have different roots.
One potential source of error is given by the time integration itself, as we discuss below.


\section{Numerical integration of time-series data}
\label{sec:integration}

The waveforms generated by numerical relativity simulations are
discretely sampled time-domain representations of a \textit{finite-length}
signal possibly contaminated by numerical ``\textit{noise}''.
The problems arising from integrating discretely sampled
numerical (or, especially, experimental) data, and are well known in other
fields of physics and engineering~\cite{Edwards05}. 
A clear analogy comes from
the use of accelerometer data to determine a position. While the
source of the problems are easy to identify, unfortunately a
rigorous solution, particularly without a detailed
characterization of the experimental noise, is difficult for
time-domain integrations.


\subsection{Time-domain integration}

Consider an integral of the form
\begin{equation}
F(t)=\int_0^t dt' f(t')\,.
\end{equation}
If $f(t)$ is known exactly, then we can evaluate the integral
numerically according to a standard scheme, and the integral will
converge to its continuum representation in the limit of infinite
resolution. However, if the function contains small amounts of experimental 
(or numerical) noise, this has a significant impact on the accuracy of the 
time integration as we will see below.

To motivate the aspect of numerical noise, consider a 
convergent finite differencing code yielding a truncation
error which can be modeled by a continuous polynomial, $p(t)$:
\begin{equation}
  f^\prime(t) = f(t) + p(t)(\Delta t)^n + O(\Delta t^{n+1}),
\end{equation}
where $f(t)$ is exact and $f^\prime(t)$ its numerical
approximation. Then, the integration yields
\begin{equation}
  \int f^\prime(t^\prime) dt^\prime = \int f(t^\prime) dt^\prime
  + (\Delta t)^n \int p(t^\prime) dt^\prime + O(\Delta t^{n+1}),
  \label{eq:convergent-integral}
\end{equation}
provided that $p(t)$ itself is sufficiently resolved.

In numerical relativity data, however, 
the convergence exponent of the measured waveforms is commonly of an
order which is higher than the lowest order of the underlying finite
differences used to generate the result, such as the
time-interpolation at mesh-refinement boundaries, or the Runge-Kutta
time integrator. That is, the results are \emph{superconvergent}. This
may come about if the error coefficient of the low order operations is
too small to be measured, or may also be associated with
under-resolution of some features of the model. Further, the
measured convergence exponents are often non-integer values, not
corresponding to the order of any discrete operation of the code,
and may vary in time, particularly during the late stages of an
inspiral simulation.
The error polynomial $p(t)$ of the one-dimensional time series
waveform is the combination of the error polynomials of a number of
independent discrete operations, including finite-differencing,
interpolation, and reduction. If any of these intermediate operations
under-resolves the cumulative error (for example, if the end product
is down-sampled), the result will be a contribution to the signal
which though deterministic, mimics the character of
numerical noise. We illustrate the effect in an analytic example.

High-frequency components of the waveforms (whether truly random
noise, a deterministic effect of discrete operators, or actual physical
modes) are aliased onto the low-frequency physical signal. This
can have a profound effect on operations such as integration.
To demonstrate this, we model the numerical estimate $g$ of some
exact function $f$ by
\begin{equation} \label{eq:function-and-noise}
  g(t_i) = f(t_i) + n(t_i)\,,
\end{equation}
where $f(t)$ is the exact result, and $n(t)$ is a continuous function
representing the truncation error of $g(t)$, sampled at discrete points
$t_i$.

\begin{figure}
  \includegraphics[width=1.\linewidth]{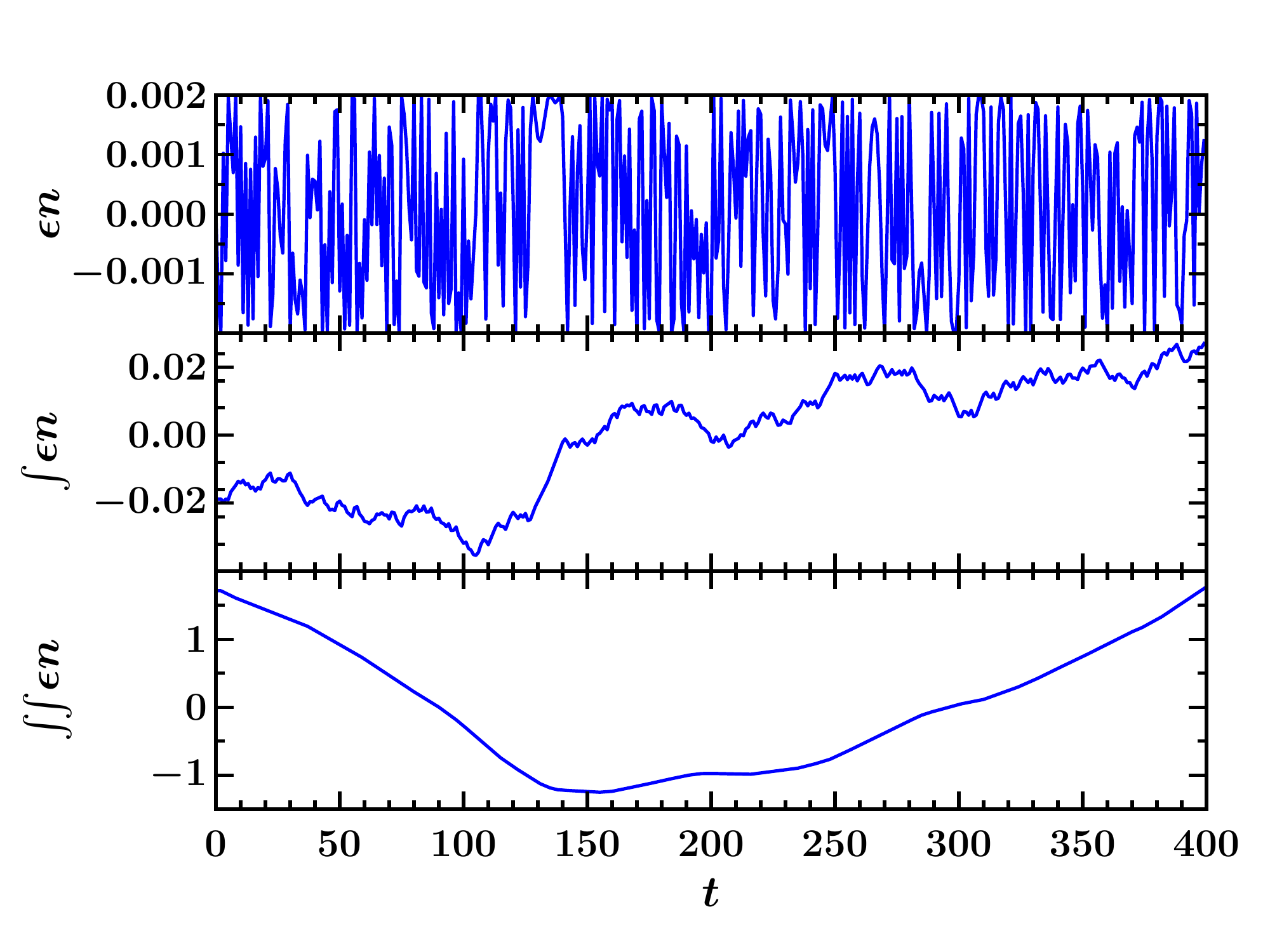}
  \caption{
    An underresolved truncation error, $n(t)$ (top
    panel) exhibits a random-walk behavior on numerical
    integration by Simpson's rule (middle panel), which in turn
    appears as a non-linear drift in the second integral (lower panel). An
    integration constant corresponding to the average of the data
    is applied after each integration to preserve the original
    oscillations about zero.
    Note that the amplitude of the effect on the second integral
    is, in this case, three orders of magnitude larger than the
    original data.}
  \label{fig:random-walk}
\end{figure}

We illustrate the effect of aliasing on integration in
Fig.~\ref{fig:random-walk}. We have integrated a signal which is
composed purely of truncation error (i.e. $f(t)=0$), modeled by
a sinusoid,
\begin{equation}
  n(t) = \epsilon \sin(\omega t),
  \label{eq:model-noise}
\end{equation}
whose frequency $\omega$ varies in time between the values $\omega_i=0.25$ and
$\omega_f=4$, according to~\eqref{eq:model-freq}, below. The upper
panel plots the original data, sampled at an interval of $\Delta t=1$.
The function $n(t)$ oscillates near the Nyquist frequency, and is
clearly under-resolved.
Its first and second integrals (in the middle
and lower panels of Fig.~\ref{fig:random-walk}, respectively) 
are analagous to what would be expected
from a random-walk. In that case, the integral over the data points
does not avarage out for large $N$.  Rather, the size of the drift,
\ie the root mean squared expected translation distance after $N$
steps, is given by the standard deviation of the imposed probability
distribution and will grow without bounds with the total number of
steps (see \eg \cite{Chandrasekhar:1943ws}).

\begin{figure}
  \includegraphics[width=1.\linewidth]{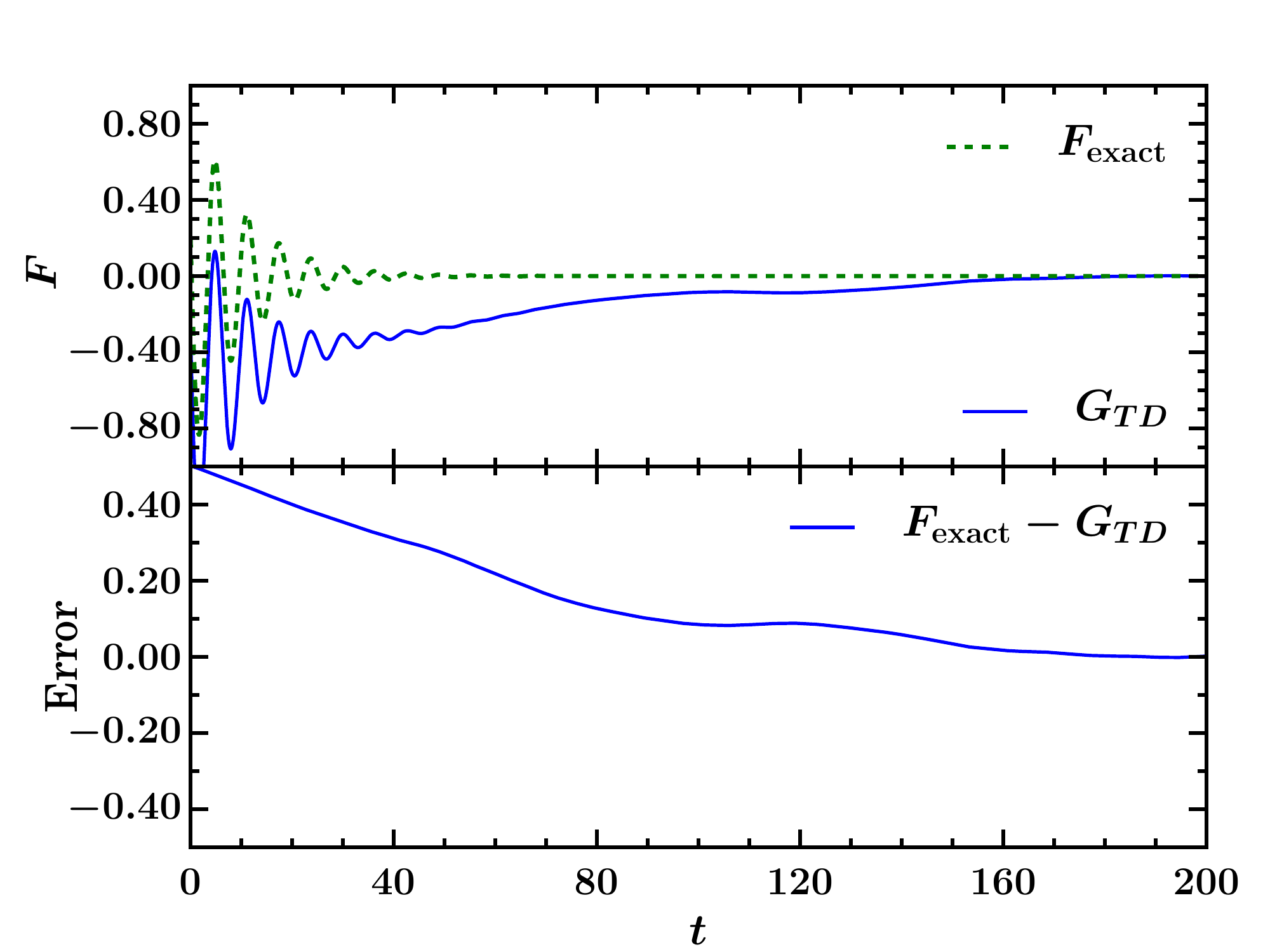}
  \caption{
    The analytic example function, \eqref{eq:example}, modified by
    an underresolved error of amplitude $\epsilon=10^{-3}$ according to \eqref{eq:function-and-noise} and \eqref{eq:model-noise}, and integrated twice in time.
    The $G_{TD}$ curve corresponds to a time-domain
    integration via Simpson's rule. Here, we have set the two integration constant
    such that the signal oscillates about zero at late times.
    There is a notable non-linear drift in the
    time-domain integration (blue solid curves).
  }
  \label{fig:example-integrated}
\end{figure}

In Fig.~\ref{fig:example-integrated}, we show the effect of an under-resolved
low-amplitude, high-frequency component on the integration of
a damped oscillatory function reminiscent of a black hole ringdown,
\begin{equation} \label{eq:example}
  f(t) = A \sin(\omega_0 t) \exp{(-t\sigma)}\,,
\end{equation}
for which we choose $A=1$ and damping parameter $\sigma=1/10$.  We fix
the frequency at $\omega_0=1$ and evaluate the function on the
interval $t\in[0,200]$, at discrete points with uniform spacing
$\Delta t = 0.05$. We directly integrate $f(t)$ numerically using a
variant of Simpson's rule (see~\cite{Press92}, for example, though as
suggested in Fig.~\ref{fig:wave_drift}, the results are largely
independent of the particular method) to compute
\begin{equation}
  F(t) = \int_0^tdt'\int_0^{t'}dt'' f(t'')\,.
\end{equation}
Since the model is analytically defined, the error in the evaluation
at each point is given by machine double-precision ($2^{-53}$, or
approximately $10^{-16}$), and the numerical integration reproduces the
exact result with high accuracy.

Consider now the effect of a small high-frequency error component
which we will again model by~\eqref{eq:model-noise}, with an amplitude
$\epsilon=10^{-3}$, and frequency parameters
$\omega_i=1/dt$, $\omega_f=1/2dt$, and $\sigma_\phi=50$.  The resulting
numerical double integration $G_{TD}(t)$, together with the analytically known double time integral $F_{\rm exact}$ of (\ref{eq:example})
is plotted in Fig.~\ref{fig:example-integrated}. A \emph{non-linear} drift,
four orders of magnitude larger than the originally induced error, is apparent in
the integral of the modified waveform.

\subsection{Frequency-domain integration of finite-length signals}
\label{sec:finite-length}

An alternative method for numerical integration of a time series
arises from simply transforming the problem to the frequency domain.
Consider the Fourier transform, $\mathcal{F}$, applied to an
absolutely integrable function $f(t)$,
\begin{equation}
  \tilde{f}(\omega) = \mathcal{F}[f] = \int_{-\infty}^\infty e^{-i\omega t} f(t) dt.
\end{equation}
The Fourier transform of the time integral of $f$ is given by
\begin{equation}
  \mathcal{F}\left[\int_{-\infty}^tdt'f(t')\right]|_\omega
    = -i\frac{\tilde{f}(\omega)}{\omega}\,,
\end{equation}  
and we arrive at a simple expression for the time-domain representation
of the integral of $f$ in terms of the inverse Fourier transform:
\begin{equation} \label{eq:Fourier-int}
  \int_{-\infty}^tdt'f(t') 
    = \mathcal{F}^{-1}\left[-i\frac{\tilde{f}(\omega)}{\omega}\right] = -\frac{i}{2\pi}\int_{-\infty}^\infty \frac{1}{\omega}e^{i\omega t} \tilde{f}(\omega) d\omega\,.
\end{equation}
In the frequency-domain, time integration becomes a simple division by
the frequency. Thus, the method is particularly susceptible to
low-frequency error. For instance, under-resolved high-frequency
modes can be aliased onto low-frequency modes of the signal.

More important is the fact
that any numerically generated (or experimentally measured) time
series is necessarily finite in length.
For frequency-domain methods, the localization of the signal in time
poses a fundamental difficulty, arising from the properties of the
Fourier transform. The observation of any finite duration signal is
equivalent of multiplying an infinite signal with a rectangular window
function.  According to the convolution theorem, multiplying a signal
with another signal in the time domain corresponds to convolving the
Fourier transformed signals in the frequency domain.  Because the
frequency representation of the rectangular window function is the
\sinc~function, which has infinite bandwidth, the same is true of the
convolved signal.

This phenomenon, sometimes termed \emph{spectral leakage}, can be
demonstrated through a simple example. Trivially, the Fourier transform of a
function of infinite extent with constant oscillation frequency, $\omega_0$,
\begin{equation} \label{eq:exp}
  f(t)=\exp(i\omega_0 t)
\end{equation}
is a Dirac delta function centred at $\omega_0$.  However, a wave of
\textit{finite} duration, can not reduce to a Dirac delta function in
the frequency domain due to spectral leakage.  Rather, the
spectrum will have a peak at $\omega_0$, but other frequencies,
particularly those close to $\omega_0$, will have non-zero values.  It
is worthwhile to emphasize that this is not an artifact of
the discrete Fourier transform.  It is an artifact of the finite
time duration of the signal.

The problem of spectral leakage has important consequences for time
integration in the frequency domain, particularly due to the division
by $\omega$ at the low-frequency end of the spectrum.  Consider the
time integral of \eqref{eq:exp} in the frequency domain, which is
trivially given by
\begin{equation}
  \int_{-\infty}^tdt'f(t') 
    \,=\, \mathcal{F}^{-1}\left[-i\frac{2\pi\delta(\omega-\omega_0)}{\omega}\right]
    \,=\, -i\frac{1}{\omega_0}\exp(i\omega_0 t)\,.
\end{equation}
If the same function is truncated in the time domain (i.e.~windowed by
the rectangular window function $\rm{rect}(t;\,t_0,t_1)$ to some
finite interval $[t_0,t_1]$), the resulting Fourier spectrum is still
peaked at $\omega_0$, but will be non-zero over an extended range. The
original delta function is ``smeared out'', and will affect the time
integral
\begin{eqnarray}\label{eq:leak}
  \int_{-\infty}^tdt'f(t')\,\rm{rect}(t;\,t_1,t_2) 
  &=& \mathcal{F}^{-1}\left[-i\frac{\mathcal{D}(\omega-\omega_0)}{\omega}
    \right]\,,
\end{eqnarray}
where $\mathcal{D}$ denotes the frequency distribution of the windowed
function arising from spectral leakage.  Division by the frequency
results in amplification of low-frequency components of
$\mathcal{D}$ other than $\omega_0$, and is responsible for secular
drifts when the time-domain signal is reconstructed.

The distribution, $\mathcal{D}$, can be modified by altering the
implicit rectangular window function associated with the finite-length
signal by tapering or fading towards the edges of the
time domain~\cite{OppenheimSchafer, Poularikas00}. However, there are
well-documented trade-offs, and the phenomenon of spectral leakage can
never be entirely compensated.

We note that for the particular case of the analytical example,
\eqref{eq:exp}, the genuine oscillation frequency is known. By
dividing only by $\omega_0$ in the function on the right-hand side
of \eqref{eq:leak}, we find that we recover in the time
domain a result which is the exact time intergal with the integration
constants set such that the signal is oscillating about zero.

In the case
of gravitational waveforms, we do not have a fixed frequency. However
for the most interesting physical models, such as late-time binary
inspiral, the range of relevant frequencies is approximately set by the initial
orbital timescale and the ringdown frequency. 
We will show in Section~\ref{sec:lbi} that exploiting this knowledge leads to an effective and
simple integration scheme which greatly reduces the impact of spectral
leakage, very similar to the simple example \eqref{eq:exp}.

\section{Optimized filters and improved frequency-domain integrations}

The effect of the spurious low-frequency modes, resulting from either
spectral leakage or aliasing effects, can be significantly suppressed
through the use of signal filters. In particular, a \emph{high-pass
  filter} can be used to reduce the energy contained in frequencies
lower than a chosen cutoff. As noted in~\cite{Santamaria:2010yb}, an
appropriate choice of filter which suppresses modes of frequency lower
than the initial instantaneous frequency of the waves, significantly
improves the form of the integral.

\subsection{High-pass filters and window functions}
\label{sec:filter}
An ideal filter is the simple step function, or a \emph{brick-wall
  filter}, which sets everything below the cut-off frequency to zero
while passing (leaving unchanged) the higher frequency components.
This filter method has previously been used
in~\cite{Campanelli:2008nk} (see also~\cite{Aylott:2009ya}) in the
context of numerical relativity waveform integration.  In practice,
however, this filter can be problematic as it gives rise to Gibbs
phenomena on transformation to the time domain. 
To suppress these effects, it is therefore preferable
to introduce a smooth transition region between the stop and pass
band.  The particular choice of the transition function is delicate,
as the wrong fall-off can result in significant oscillations in the
amplitude of the reconstructed time-domain signal.

The tapering function (or \emph{window}) is effectively the transfer
function, $H(\omega)$, defined by
\begin{equation}
  H(\omega)=\frac{Y(\omega)}{X(\omega)}\,,
\end{equation}
where $X(\omega)$ is the original signal, and $Y(\omega)$ the filtered
signal. Applying a window function is equivalent to imposing a certain
function to $H(\omega)$ in order to arrive at the filtered signal
from the original data.

Santamaria et al.~\cite{Santamaria:2010yb} apply a $\tanh$ window of the form:
\begin{equation} \label{eq:tanh-window}
  H(\omega) = \frac{1}{2} \left[ 1 + 
    \tanh\left(\frac{4(\omega - \omega_0)}{\sigma}\right)\right]
\end{equation}
where the parameters $\omega_0$, $\sigma$ must be chosen. Meanwhile,
McKechan et al.~\cite{McKechan:2010kp} have analyzed the properties of
a tapering window in the time domain based on a Planck-distribution,
in order to minimize oscillations in the frequency domain.  In
practice, applying a transfer function such as
\eqref{eq:tanh-window} in the frequency domain requires some
non-trivial fine-tuning as there are two free parameters, $\omega_0$
and $\sigma$, which have a sensitive effect on the removal of non-linear drifts
in the reconstructed time-domain signal. 
In addition, the choice of window parameters are not easily 
generalizable to different simulation models and the various other (higher) 
harmonic modes as each mode 
requires an indiviudal and different set of fine-tuned parameters.

To circumvent these problems, we propose a different \emph{ansatz}.
Instead of imposing a particular transfer function through
the choice of a fixed window function,
we derive an appropriate transfer function,
$H(\omega)$, from the data in order to reduce the amount of required
fine-tuning significantly. Our proposal is guided by the
following observation.  By setting the power spectrum to
\begin{equation} \label{eq:butterworth}
 |\tilde{f}(\omega)| = a\,\omega^b,\qquad \omega \leq \omega_0\,,
\end{equation}
for frequencies below a chosen frequency $\omega_0$, we find that we
can minimize non-linear drifts arising in the time domain by an appropriate
choice of the parameters $a$ and $b$, and
this behavior is generic for different (higher) harmonic modes.
This frequency fall-off
is similar to that of a Butterworth filter~\cite{butterworth:1930,
  OppenheimSchafer}, known to result in a maximally flat response in
the frequencies that are passed. The empirically observed result of
applying such a filter is to suppress non-linear drifts of the centre of the
waves away from zero, which is the source of
ripples in the amplitude.  The drawback is a slower roll-off towards
low frequencies, which means that part of the signal at low
frequencies will be lost due to the transition.

\begin{figure}
  \includegraphics[width=1.\linewidth]{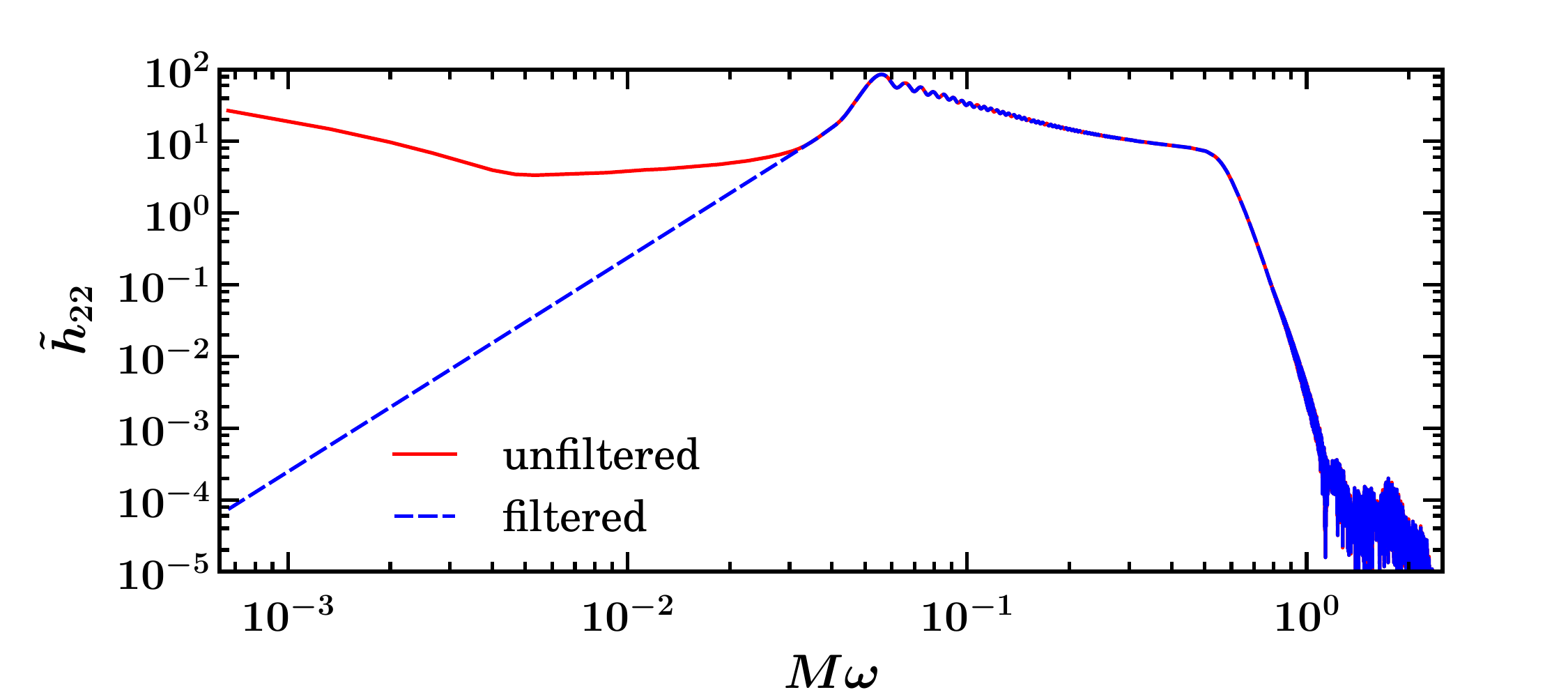}
  \caption{
    Power spectrum $|\tilde{h}_{22}(\omega)|$ in the frequency domain of an
    equal-mass non-spinning binary black hole merger simulation.
    In the blue curve the low frequencies are significantly amplified
    due to the division by very small numbers, $\omega\ll1$.  The
    `filtered' curve (blue, dashed), on the other hand has low
    frequency components determined by a polynomial of the form
    $y=ax^b$ for frequencies below the initial instantaneous frequency.
    This choice limits the spurious frequency oscillations in the
    time domain. The same low-frequency fall-off can principally also be achieved through a window
    as given by (\ref{eq:tanh-window}), however, not without a certain amount of fine-tuning.
    The plotted waveform has an initial instantaneous frequency $\omega_{\rm i}=0.05$
    and we have used $\omega_0=0.034$ and $\omega_1=0.035$ for the filter settings.
    The fitting coefficients become $a=2.0754\times10^9$ and $b=4.9823$.
  }
  \label{fig:h-filtered-vs-unfiltered-fourier}
  \includegraphics[width=1.\linewidth]{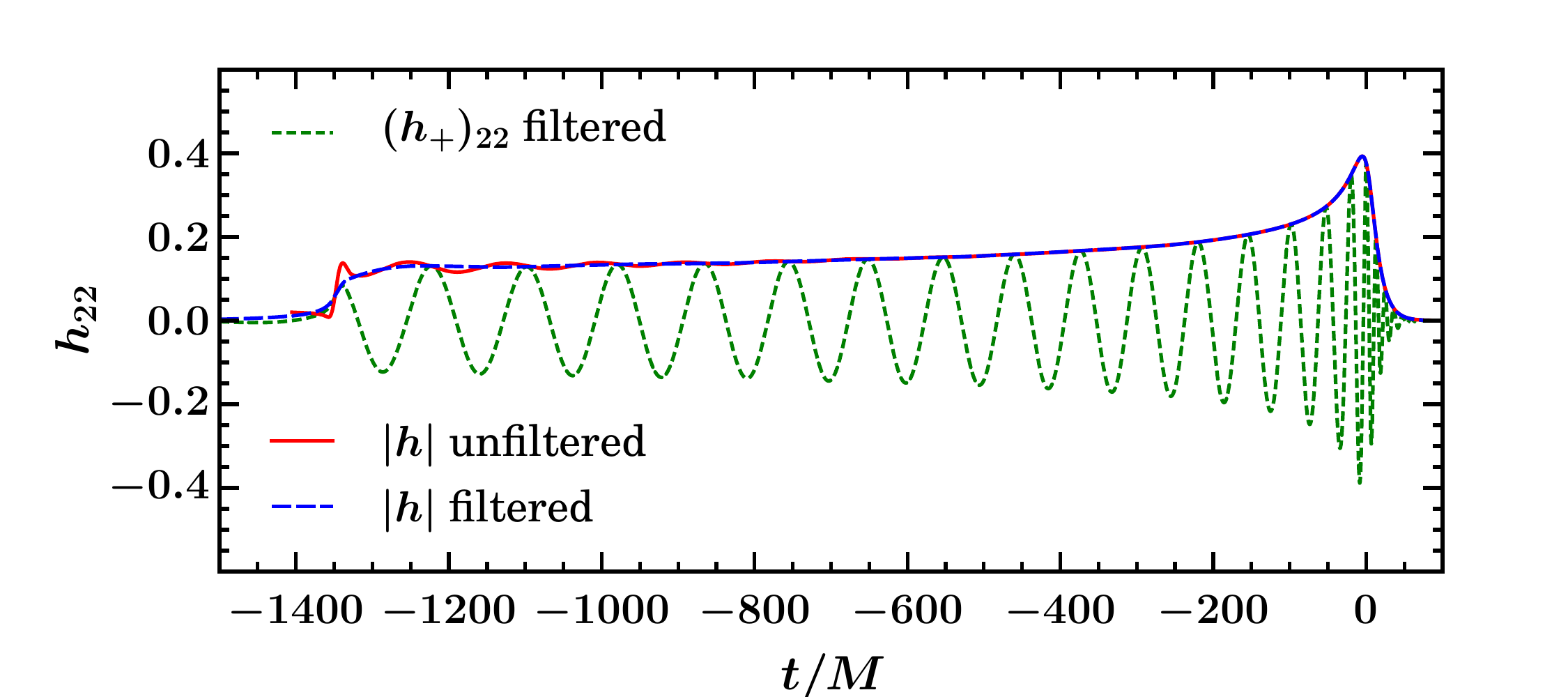}
  \caption{
    The twice integrated $(\psi_4)_{22}$ wavemode of an equal-mass non-spinning binary black hole merger.
    The wavemode is the same that was used to compute $h$ and its spectrum in the figure above.
    We choose the integration constants such that after each integration, the signal is zero at late times.
    The integrated unfiltered signal exhibits spurious oscillations in
    the complex amplitude, $|h|$ (red, solid line), as a result of a non-linear
    drift in the circularly polarized wave. With a careful choice of
    filter parameters (see text), the filtered integration essentially
    is essentially free of non-linear drifts. We plot its amplitude (blue, long
    dashed) and the $h_+$ component (green, short dashed).
  }
  \label{fig:h-filtered}
\end{figure}

Specifically, we can carry out integrations as follows. First, we
transform individual oscillating $(\ell, m)$ spherical harmonic modes
of $\psi_4$ to the frequency domain. In a $\log$-$\log$ plot, functions
of the form \eqref{eq:butterworth} are linear, with slope
$b$. Empirically, we find that we can remove the drifts in the
time domain by setting the power spectrum to
\begin{equation}
  |\tilde{\psi}_4|=a\omega^b\,,\qquad a, b\in\mathbb{R},
\end{equation}
where the coefficients $a$ and $b$ are determined by fitting to a
section of the waveform over a chosen interval $[\omega_0, \omega_1]$,
which is below the lowest instantaneous physical frequency $\omega_{\rm i}$ 
of the model, and where the spectrum is approximately linear in
the $\log$-$\log$ plot. We compute
\begin{equation}
  b = \frac{\log |\tilde\psi_4(\omega_1)|/|\tilde\psi_4(\omega_0)|}{\log (\omega_1/\omega_0)}\,,
     \qquad \omega_0<\omega_1\,.
\end{equation}
(where `$|$' indicates the complex modulus). From this we determine
\begin{equation}
  a = \frac{|\tilde\psi_4(\omega_1)|}{\omega_1^b}\,.
\end{equation}
In order to control the power spectrum below $\omega_1$, we compute the
transfer function according to
\begin{equation}
  H(\omega) = \frac{a\omega^b}{|\tilde\psi_4(\omega)|}\,,
\end{equation}
and use pointwise multiplication to determine
\begin{equation} \label{eq:psi4_filtered}
  \tilde\psi_4^{\rm filtered}(\omega)
   = H(\omega)\cdot\tilde\psi_4(\omega)\,, \qquad 0\leq\omega\leq\omega_1\,.
\end{equation}
We can then determine $h$ as the 2nd time integral of $\psi_4$,
according to Eqs.(\ref{eq:h_from_N}) and (\ref{eq:Fourier-int}),
\begin{equation}
  \th(\omega)=-\frac{\tilde\psi_4^{\rm filtered}(\omega)}{\omega^2}\,,
\end{equation}
and apply the inverse Fourier transform to obtain the result in the
time domain.  With a careful choice of a frequency fitting interval, $[\omega_0, \omega_1]$
we find that the resulting strain, $h(t)$, is free of non-linear drifts and
spurious oscillations.  An example is plotted in
Figs.~\ref{fig:h-filtered-vs-unfiltered-fourier}
and~\ref{fig:h-filtered}.
Once an optimal set of parameters $\omega_0$, $\omega_1$ has been found,
the same procedure can be applied to higher harmonic modes without additional fine-tuning
by using the relation $\omega_{\ell m}=m\omega_{22}/2$, which is a result of
the phase relation of the spherical harmonics ${}_sY_{\ell m}$.

Although the filtering technique is effective in limiting non-linear drift
artifacts of the reconstructed time-domain signal, a principle
drawback is the assumption that the power spectrum of the transformed
signal contains a segment below $\omega_{\rm i}$ which allows for a
linear fit in the log-log function plot. While empirically this is the case for binary black hole
waveforms, we find that more general waveforms such as signals from stellar core collapse
signals (see \eg \cite{Ott09}) do not have such simple power spectra, making the choice of falloff exponent, $b$, more difficult.

\subsection{Fixed frequency integration (FFI)}
\label{sec:lbi}

A much simpler, yet empirically more robust, method for integrating in
the frequency domain is suggested by the simple example discussed in
Sec.~\ref{sec:finite-length}. For the analytically known function
with a single frequency component, the integration is greatly
improved (in the sense of removing spurious non-linear drifts) by
applying information about the expected frequency band to control the
amplification of the unphysical frequencies resulting from spectral
leakage.  That example involves only a single oscillation frequency,
$\omega_0$, and by simply multiplying with $-i/\omega_0$ it is
possible to achieve drift-free integration.

However, an astrophysically interesting waveform such as that of a
binary black hole merger is characterized by a range of physical
frequencies, primarily determined by the intial orbital velocity
$\omega_i$ and increasing to the ringdown frequency, $\omega_{\rm
  QNM}$ for each $(\ell,m)$ mode, so that the true physical frequency
content is approximately\footnote{We note that the true physical frequency 
range is slightly larger. For instance, the exponentially damped ring-down signal 
is a Lorentz distribution in the Fourier domain, even though it contains a single QNM oscillation frequency. 
For exponentially damped signals, a more natural way of describing the frequency content is given by the Laplace transform
where damped signals with a single (complex) frequency transform to a Dirac delta function.} 
within $\omega\in[\omega_i,\omega_{\rm QNM}]$.  In the
Fourier transformed wave, any frequencies outside of this band are
dominated by the effect of spectral leakage of the finite length
time-domain signal.

By effectively applying the same integration method as proposed for the example
\eqref{eq:exp} to the range of physically relevant
frequencies, $\omega\in[\omega_i,\omega_{\rm QNM}]$, we find that the
spurious non-linear drifts due to the amplification of unphysical
(spectrally-leaked) low-frequency modes are essentially removed.

Accordingly, we propose to evaluate the integral using the following prescription:
\begin{equation}\label{eq:int-scheme}
   \tilde{F}(\omega) = \left\{
     \begin{array}{lr}
       -i\tilde{f}(\omega)/\omega_0\,, & \omega \leq \omega_0\,, \\
       -i\tilde{f}(\omega)/\omega\,, & \omega > \omega_0\,.
     \end{array}
   \right.
\end{equation}
In order to get the second integral, we simply divide by $(-i\tilde{f}(\omega)/\omega_0)^2$
and $(-i\tilde{f}(\omega)/\omega)^2$, respectively.
The single free parameter is $\omega_0$, which is set according to the
lowest expected physical frequency for the given wave mode. When
adjusted correctly, we find that the time-domain representation of the
waveform is essentially free of low-frequency drifts.
An upper, high-frequency, integration limit via some additional parameter 
$\omega_1 > \omega_{\rm QNM} > \omega_0$ could also be incorporated
in \eqref{eq:int-scheme}, however is not needed for the
particular waveforms studied here, given the exponential ringdown,
combined with the fact that high frequency errors are not as strongly
amplified on integration.

In practice, the choice of $\omega_0$ requires a certain amount of
tuning. A small value will amplify unphysical low-frequency components
during the integration process, while a large value may suppress some
desired physical frequencies of the waveform. However, the choice is
clearly guided by the known features of the original signal, and as
will be demonstrated below, improved integrations result from a broad
range of the choice of $\omega_0$.

The main advantage of the FFI method over windowing
functions as described in the previous section is simplicity and
generality. By tuning a single parameter, we are able to eliminate the
bulk of the linear and non-linear drift in the waveform. This frees the integration from
ambiguities in the choice of optimal windowing functions and their
respective parameters and can easily be automatized for all higher modes
through the relation $\omega_{\ell m}=m\omega_{22}/2$,
once a particular $\omega_0$ has been found for the dominant $(\ell,m)=(2,2)$
mode.

As a final remark, we note that in certain situations, the result
of the FFI can be improved by first applying a window function 
to the time domain signal before transforming to the Fourier domain.
This is particularly the case for signals which do not start and end with zero amplitude.
In these situation, a window function of the form \eqref{eq:tanh-window}
may be applied such that the signal smoothly blends from and to zero at beginning and end, respectively. 
Emperically, we find that the choice of time-window parameters does not require much fine-tuning as
long as the transition region is chosen to be sufficiently broad.

\subsection{Error estimates for an analytic model}
\label{sec:est-error}

The frequency-domain integration methods avoid the random-walk effects
associated with time-domain integration, however they are only able to
reduce the problem of spectral leakage at the cost of modifying the
original data by introducing spurious low frequencies (via the Fourier
transform) which must subsequently be suppressed. A concern is that in the process, physical information
may be lost or altered.  The magnitude of this effect is difficult
to gauge using numerical waveform data for which an exact solution for
$h(t)$ is not known \emph{a priori}. We apply the method to an
analytic model which exhibits the main features of a binary chirp
signal, so that the effects of numerical error and integration methods
can be compared against a known result.

\begin{figure}
  \includegraphics[width=\textwidth]{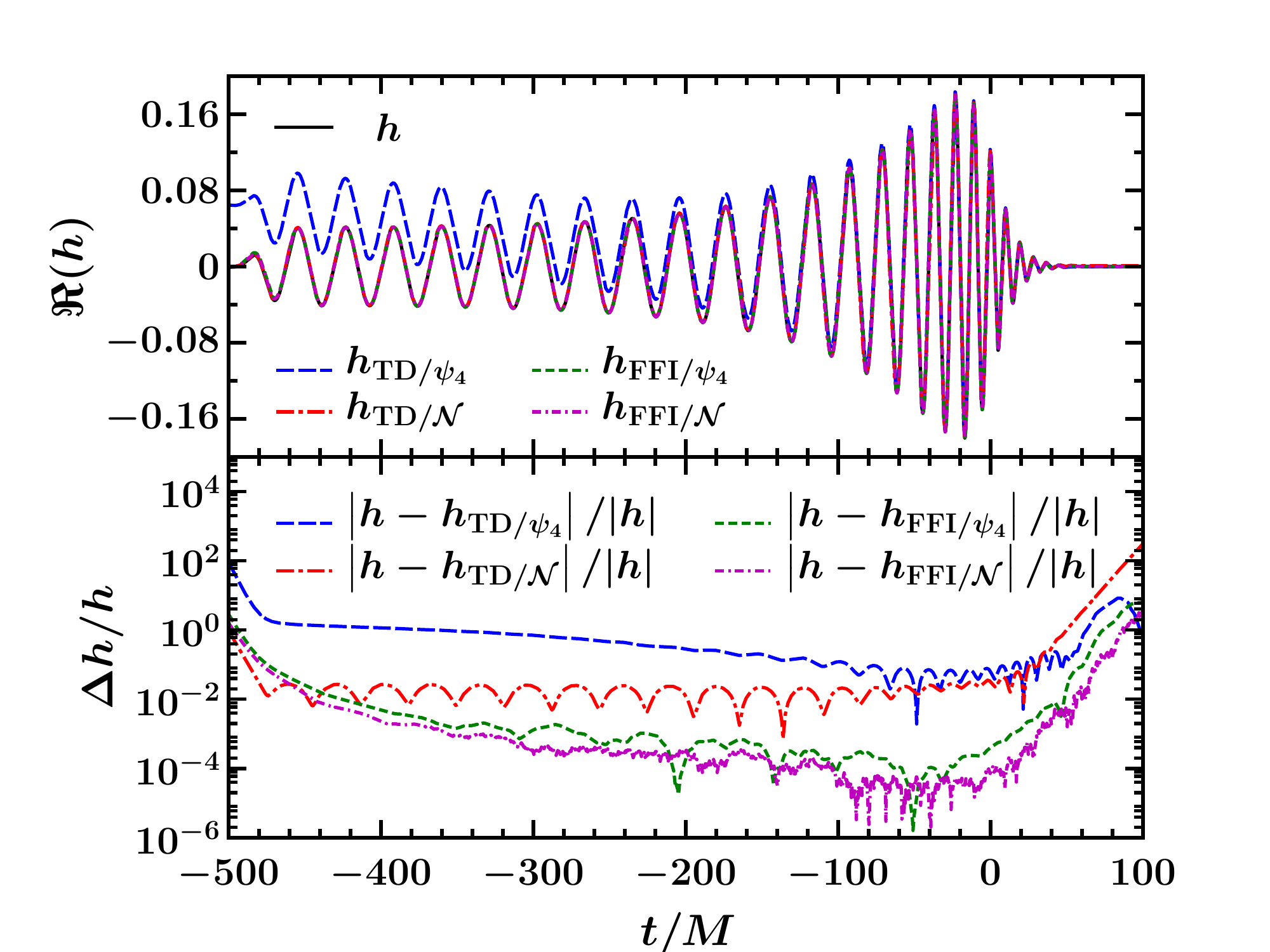}
  \caption{
    A comparison of time-domain integration with FFI, using the
    analytic function~\eqref{eq:example_analytic_wave}. For each case,
    the news, \news, or $\psi_4$ are determined analytically at
    uniformly sampled points ($dt=0.1$) and then adjusted by
    underresolved error of amplitude $10^{-5}$. Time-domain
    integration is used to determine the curves $h_{\rm TD/\psi_4}$
    and $h_{\rm TD/\news}$ from $\psi_4$ and \news, respectively. 
    The integration constants are choosen such that the signal is zero at late times.
    The FFI method with $\omega_0=0.15$ is used to determine $h_{\rm
      FFI/\psi_4}$ and $h_{\rm FFI/\news}$. Similar results to FFI can
    be achieved with the filter methods as discussed in Sec.~\ref{sec:filter},
    though with some tuning required.
  }
  \label{fig:analytic_example}
\end{figure}

We introduce a simple analytic toy-model which provides a rough
approximation to some of the properties of a typical inspiral waveform
over some cycles, including the merger and ringdown.  We construct an
artificial strain according to the oscillating function
\begin{equation} \label{eq:example_analytic_wave}
  h(t) = A(t)\, e^{-i\, \phi(t)}\,,
\end{equation}
where
\begin{eqnarray}
  \fl A(t) & = & \frac{A_1}{2}
    \left[1 + \tanh\left(\frac{t-t_0}{\sigma_0}\right)\right] \times 
     \left(1 + A_2\exp\left(\frac{t}{\sigma_2}\right) 
     \left[1 + \tanh\left(\frac{-(t-t_1)}{\sigma_1}\right)\right]
     \right), \\
  \fl \phi(t) & = & \omega_{\rm i} (t-t_1) + \left(
    \frac{\omega_{\rm f} - \omega_{\rm i}}{2}\right)
    \left[1 + \sigma_\phi \log\cosh\left(\frac{t-t_1}{\sigma_\phi}\right)
    \right]\,.
    \label{eq:model-phase}
\end{eqnarray}
Here, $\tanh$ functions have been used to control various
transitions between essentially constant values. The amplitude $A(t)$ rises from zero
at time $t_0$ over a distance $\sigma_0$, to an amplitude of approximately
$A_1$. The choice of parameters $A_2$, $\sigma_1$, $\sigma_2$, and
$t_1$, control the size, shape and location of an eventual peak in
the amplitude. The choice of phase, $\phi(t)$, leads
to a frequency evolution of the form
\begin{eqnarray}
  \omega(t) & = & \omega_{\rm i} + \left(
     \frac{\omega_{\rm f} - \omega_{\rm i}}{2}\right) 
     \left[1 + \tanh\left(\frac{t-t_1}{\sigma_\phi}\right)\right]\,. \label{eq:model-freq}
\end{eqnarray}
The frequency transitions from an initial value of approximately
$\omega_{\rm i}$ for small $t$, to $\omega_{\rm f}$ as $t\rightarrow +\infty$,
over an interval whose location and width are controlled by $t_1$ and
$\sigma_\phi$, respectively.  An example profile
for~\eqref{eq:example_analytic_wave} is shown in the upper panel of
Fig.~\ref{fig:analytic_example}, corresponding to the particular
parameter choices:
\numparts
\begin{eqnarray}
  \{t_0\,, t_1\} & = & \{-480.0\,, 0.0\}\,, \\
  \{\omega_{\rm i}\,, \omega_{\rm f}\} & = & \{ 0.2\,, 1.0 \}\,, \\
  \{A_1\,, A_2\} & = & \{ 0.02\,, 5.0\}\,, \\
  \{\sigma_0\,, \sigma_1\,, \sigma_2, \sigma_\phi\} & = &
  \{10.0\,, 16.0\,, 80.0\,, 80.0\}\,.
\end{eqnarray}
\endnumparts

\begin{figure}
  \includegraphics[width=\textwidth]{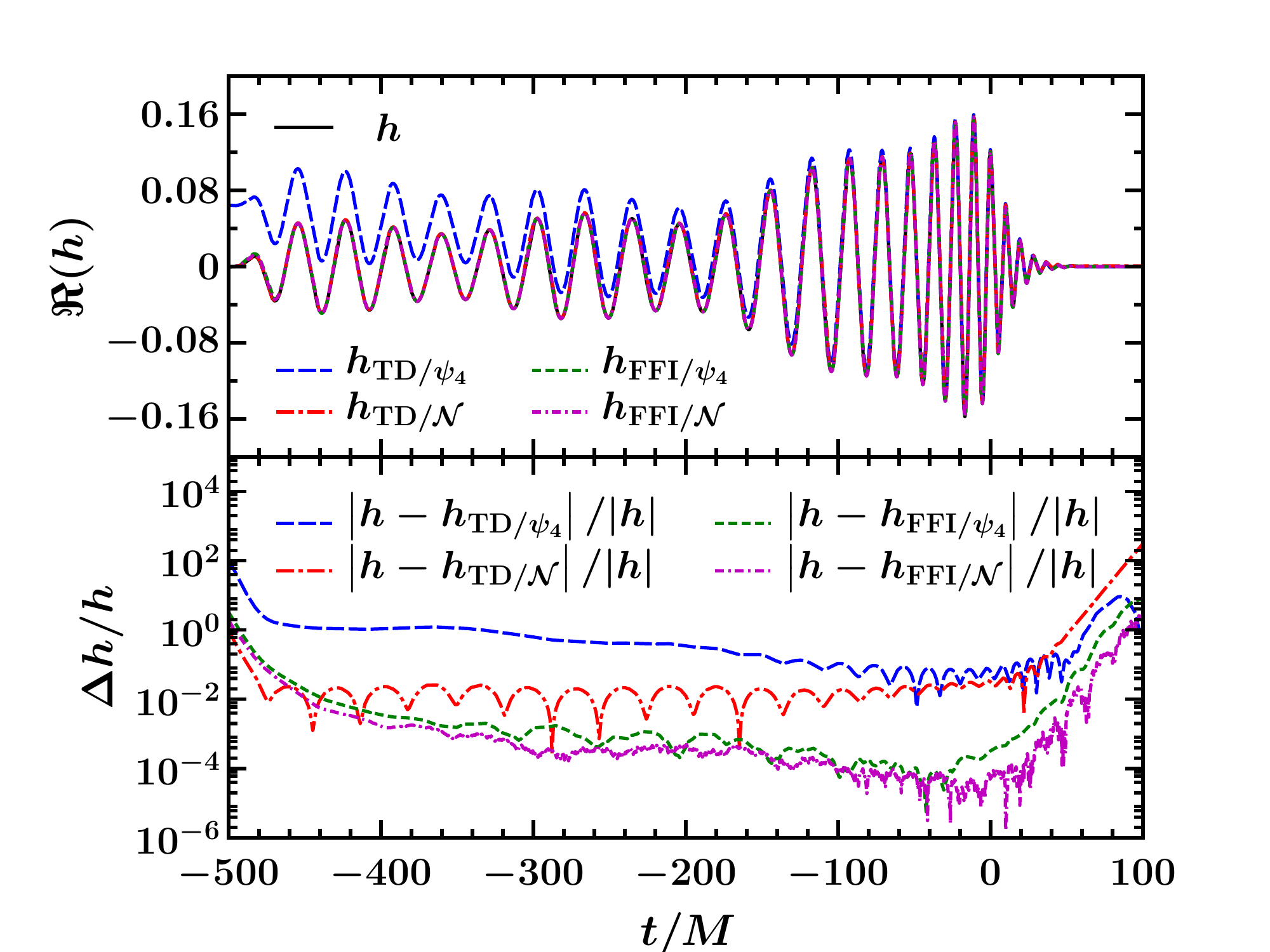}
  \caption{
    A comparison of time-domain integration with FFI, using the
    analytic function~\eqref{eq:example_analytic_wave} where an
    additional low-frequency modulation has been applied, according
    to~\eqref{eq:modulation}.  Curves are determined as described in
    Fig.~\ref{fig:analytic_example}.
  }
  \label{fig:analytic_example_modulated}
\end{figure}

We test the numerical integration methods by determining
analytic news, \news, and curvature component $\psi_4$, functions
according to
\begin{equation}
  \news = dh/dt\,, \qquad \psi_4 = d^2h/dt^2\,.
\end{equation}
These functions are sampled at discrete points, $i$, over an interval,
and adjusted by the sinusoidal error function,
Eq.~\ref{eq:model-noise}, to simulate a level of underresolved
truncation error in the numerical data:
\begin{equation}\label{eq:added_noise}
  \news \rightarrow \news + \epsilon n_i\,, \qquad
  \psi_4 \rightarrow \psi_4 + \epsilon n_i\,.
\end{equation}
We then reconstruct $h$ by performing numerical integrations of
\eqref{eq:added_noise}, and compare the result with the original
analytic function, \eqref{eq:example_analytic_wave}.

Some representative results are plotted in
Fig.~\ref{fig:analytic_example}.  For this test, we have sampled
$\news(t)$ and $\psi_4(t)$ at 6000 equally spaced points ($dt=0.1$)
over an interval from $t=-500$ to $t=100$, and adjusted the data by an
error signal modeled by the underresolved wave
Eq.~\ref{eq:model-noise} with an amplitude $\epsilon=10^{-5}$.  We
compare the strain computed by performing the integrals of \news and
$\psi_4$ in the time domain (TD, via a 4th-order Simpson's rule), with
those of the FFI method, described in the previous section. For the
latter, we note that the starting frequency for the test waveform is
$\omega_{\rm i}=0.2$, and thus choose a somewhat smaller cutoff
frequency $\omega_0=0.15$ for the FFI scheme described by
\eqref{eq:int-scheme}.  The results show a prominent drift for the
case of two time-domain integrations ($h_{TD/\psi_4}$). The situation
is greatly improved if only a single integration is required
($h_{TD/\news}$). The lowest level of error results from the FFIs,
again with a slight advantage if only a single integration needs to be
performed. The results are robust against the particular
choice of $\omega_0$, and we find that values between 0.02 and 0.2
outperform the time-domain integration at this level of error. Similar
levels of error can also be attained by the high-pass filter methods,
described in Sec.~\ref{sec:filter}, with correctly chosen parameters.

A particular concern with the FFI method (as well as with the use of
high-pass filters, as in the previous section) is that any genuine
physical low-frequency information will be modified during the
integration process. In a binary system, lower frequency components
may arise, for instance, due to precession effects, or ellipticity
(including zoom-whirl behaviour \cite{Pretorius:2007jn, Sperhake:2009jz}).  
We mimic the presence of such
features in the toy-model by modulating the amplitude of the analytic
wave according to the function
\begin{equation}\label{eq:modulation}
  \tilde{A}(t) = \left[1 + A_m \sin(\omega_mt)\right] A(t),
\end{equation}
where $A_m$ determines the amplitude of the new component, and
$\omega_m$ its frequency. In
Fig.~\ref{fig:analytic_example_modulated}, we plot the analytic $h(t)$
determined by \eqref{eq:example_analytic_wave} with $A(t)$ replaced by
$\tilde{A}(t)$, using the parameters $A_m=0.2$,
$\omega_m=0.2\omega_0$. We find that since the low-frequency mode
has a rather small influence on the Fourier spectrum, the
FFI method accurately reproduces the mode in the integrated waveform
and continues to outperform time-domain integration.
A possible explanation for this behavior might be given by the fact that
an amplitude modulation like \eqref{eq:modulation} results in
additional effective oscillations (sidebands) of frequency $\omega_{\rm eff}=\omega\pm\omega_m$
in the signal. Thus, if $\omega_m$ is small, then the effective contributing 
lowest frequency $\omega_{\rm eff} = \omega_{\rm i}-\omega_m$
is only slightly lower than the initial orbital frequency $\omega_i$.

It is difficult to make rigorous quantitative statements about the
expected level of error based on these tests, particularly since the
analytic waveform is only superficially similar to a genuine inspiral
model. Tests with a variety of alternate functions and
parameter choices, however, suggest that the qualitative picture is robust.  The
FFI method provides a reliable means to reduce integration error over
time-domain integration at a given level of numerical error. It is not
surprising that it is generally preferable to perform a single
integration rather than two. Thus, if the strain $h$ is the desired
product, then raw numerical data in the form of the news, \news, or
Zerilli-Moncrief variables, have an advantage. And finally, regardless
of the integration method used, it seems to be difficult to reliably
estimate amplitude errors to within $\simeq 1\%$ if an exact target
solution is not known \emph{a priori}.

\subsection{Application to numerical waveforms}
\begin{figure}
  \includegraphics[width=1.\linewidth]{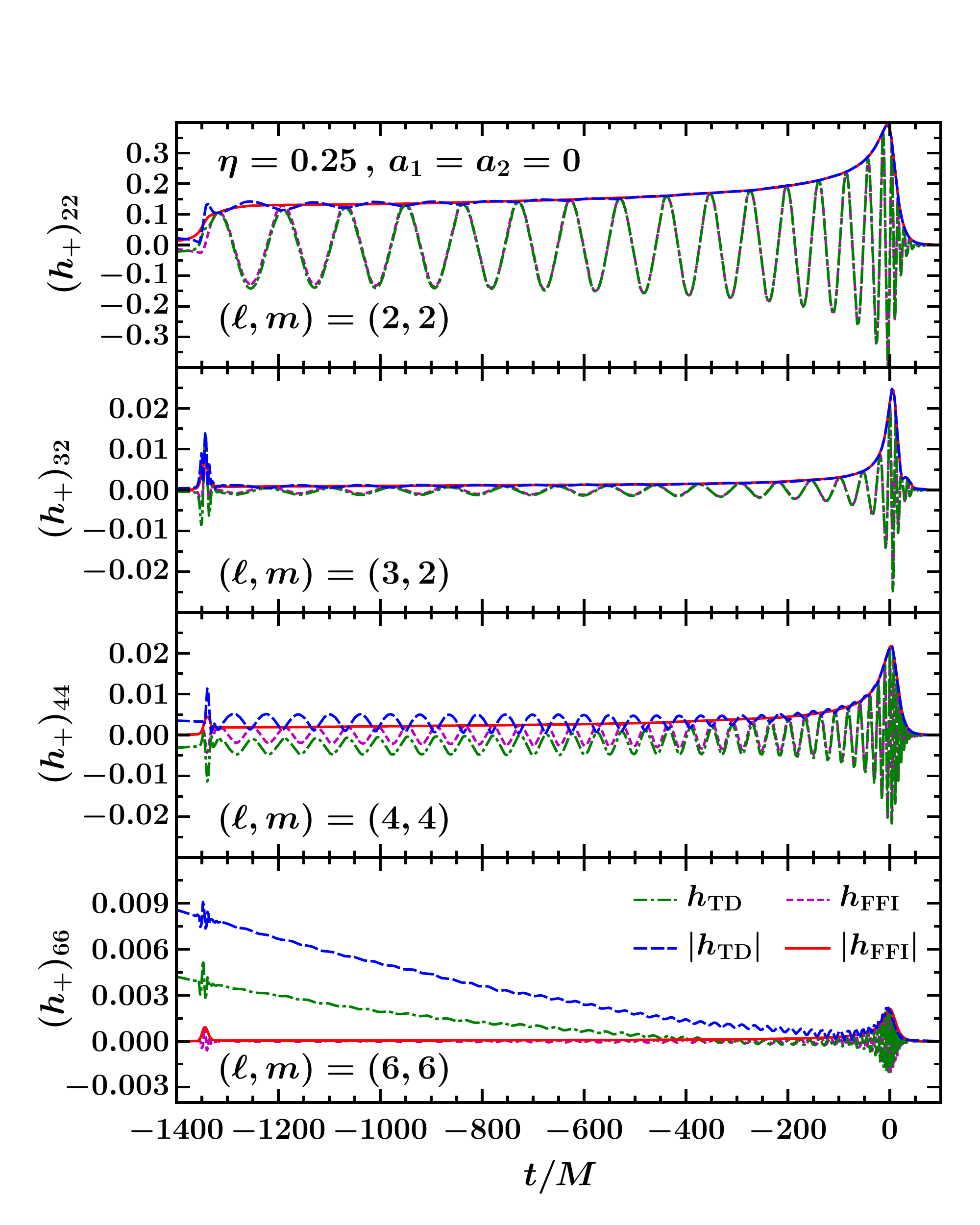}
  \caption{
    Various modes of the gravitational-wave strain component $h_+$, for an equal-mass, $\eta=0.25$,
    non-spinning, $a_1=a_2=0$, binary, computed from $\psi_4$ via standard
    time-domain integration, and via FFI in
    the frequency domain. 
    The integration constants are choosen such that the signal is zero at late times.
    From top to bottom, the $(\ell,m)=(2,2)$,
    $(3,2)$, $(4,4)$ and $(6,6)$ modes are plotted, respectively.
    Time-domain integration generically exhibits a notable non-linear drift from
    zero, visible in the oscillations of the wave amplitude $|h|$. A
    simple frequency-domain integration via~\eqref{eq:Fourier-int}
    results in drifts which are off the scale on these axes.  The drifts are greatly
    suppressed through the FFI method, as can be seen from the
    non-oscillatory (red, solid) line in each panel.
  }
  \label{fig:psi4-integrated-d550}
\end{figure}

\begin{figure}
  \includegraphics[width=1.\linewidth]{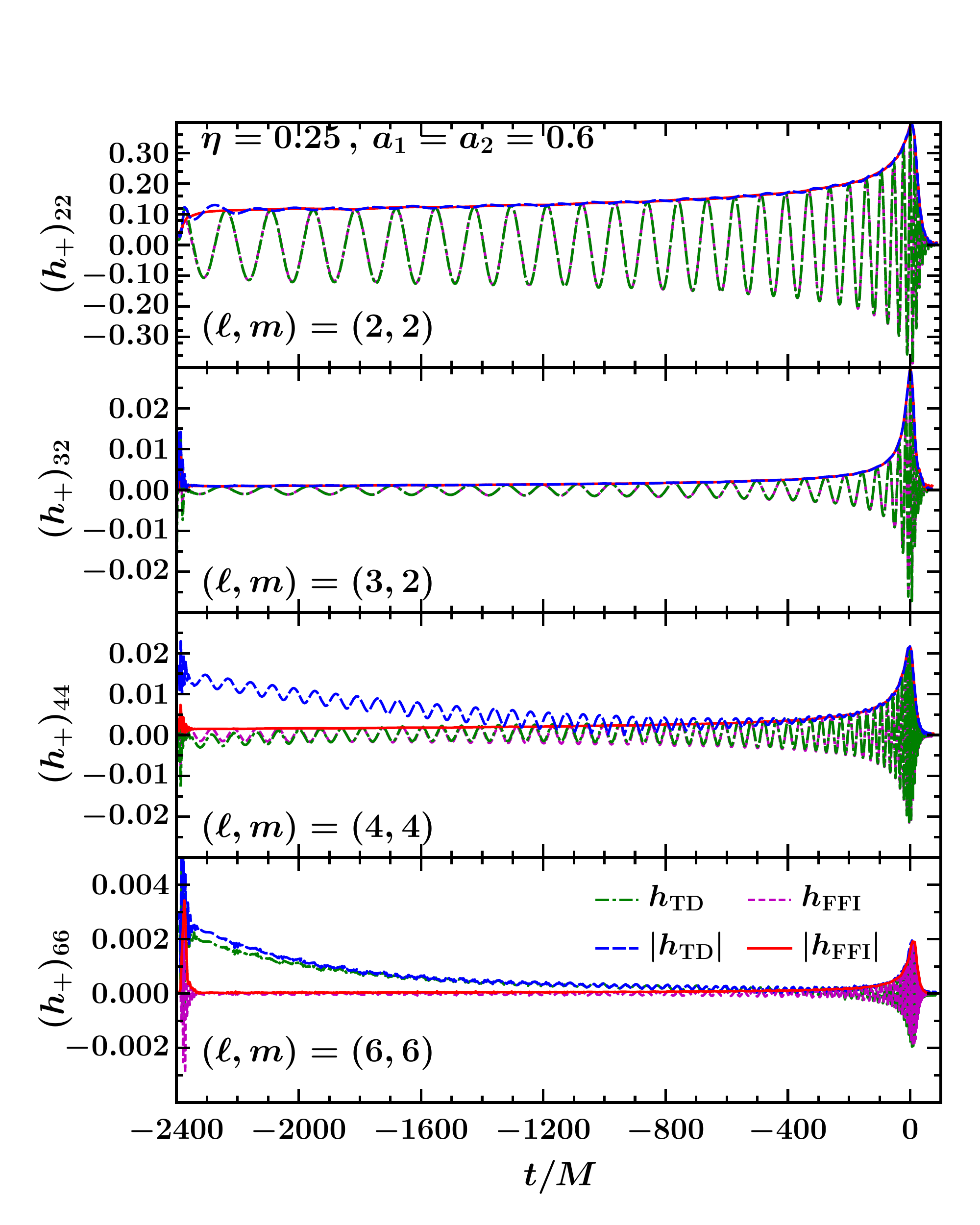}
  \caption{Various modes of the gravitational-wave strain component $h_+$, for an
    equal-mass, $\eta=0.25$, binary for which each body has spin, $a_1=a_2=+0.6$,
    aligned with the orbital angular momentum. The same analysis as in Fig.~\ref{fig:psi4-integrated-d550} applies.}
  \label{fig:psi4-integrated-uu}
\end{figure}

The quality of various integration schemes can be readily seen, if not
precisely quantified, in the results of numerically generated
waveforms. We present plots from two example models in
Figs.~\ref{fig:psi4-integrated-d550} and~\ref{fig:psi4-integrated-uu}.

The first of these is a non-spinning equal mass binary, presented
in~\cite{Pollney:2009ut, Pollney:2009yz, Reisswig:2009us,
  Reisswig:2009rx}.  We plot four different spherical harmonic modes
of the strain, $h(t)$, calculated by integrating $\psi_4$ which was
evaluated during the simulation at future null infinity, $\scri$. The
time series for $\psi_4$ has a resolution of $dt=0.144$. 
In this case, the truncation error appears as high frequency error, effectively resulting 
in a numerical truncation error of approximately $10^{-6}$. For this model, 
we have choosen the integration constants such that the signal becomes zero at late times.
Hence, each of the
displayed modes is expected to oscillate about zero. However, we
notice a slight non-linear drift in the time-domain integrated $h_{\rm
  TD}$, computed using a 4th-order Simpson's rule. The effect of the
drift is clearest in the plot of the complex amplitude, $|h|$, which
should grow monotonically, but rather displays oscillations at half
the orbital frequency wherever the circularly polarized modes are
off-centred. We have integrated the same $\psi_4$ data using the FFI
method. The initial orbital frequency at the start time of the
simulation is $\omega\simeq 0.025$, corresponding to a wave
frequency of $\omega_{\rm i}=0.05$ in the $(\ell,m)=(2,2)$
mode. For the integration procedure, we have used $\omega_0=0.035$
for the $(2,2)$ and $(3,2)$ modes, and $2\omega_0$ and $3\omega_0$ for
$(4,4)$ and $(6,6)$ respectively. The resulting strain shows that
drifts have been strongly reduced, while maintaining the overall wave
amplitude. (The latter point can be gauged approximately by the fact
that the FFI amplitude tracks the average of the oscillations of the
TD amplitude in the $(2,2)$ case, or by shifting the waves and
comparing the amplitudes of individual cycles.)

Similar results are apparent in Fig.~\ref{fig:psi4-integrated-uu}. In
this case, the waveform is from a model for which each body has spin
$+0.6$ aligned with the orbital angular momentum. The sampling rate
and truncation error are the same as in the previous model. In this case,
the initial frequency of the $(2,2)$ mode is
$\omega_{\rm i}=0.044$.  For the FFI method, we have used
the same $\omega_0$ as in the non-spinning case, applied to each
mode. The integration is most sensitive to the choice of $\omega_0$ in
the early part (first $200M$) of the wave, and late ringdown ($t>50M$ after the peak) when
the amplitude approaches the level of the truncation error. By varying the
integration parameter between $\omega_{\rm i}/2\leq \omega_0\leq
\omega_{\rm i}$, we find variations of $8\%$ and $50\%$ in the
calculated amplitude in these two regions, respectively. However,
restricting attention to the range $t\in[-2000,40]$, we find that
varying the integration parameter affects the calculated amplitude by
less than $1\%$.

\begin{figure}
  \includegraphics[width=1.\linewidth]{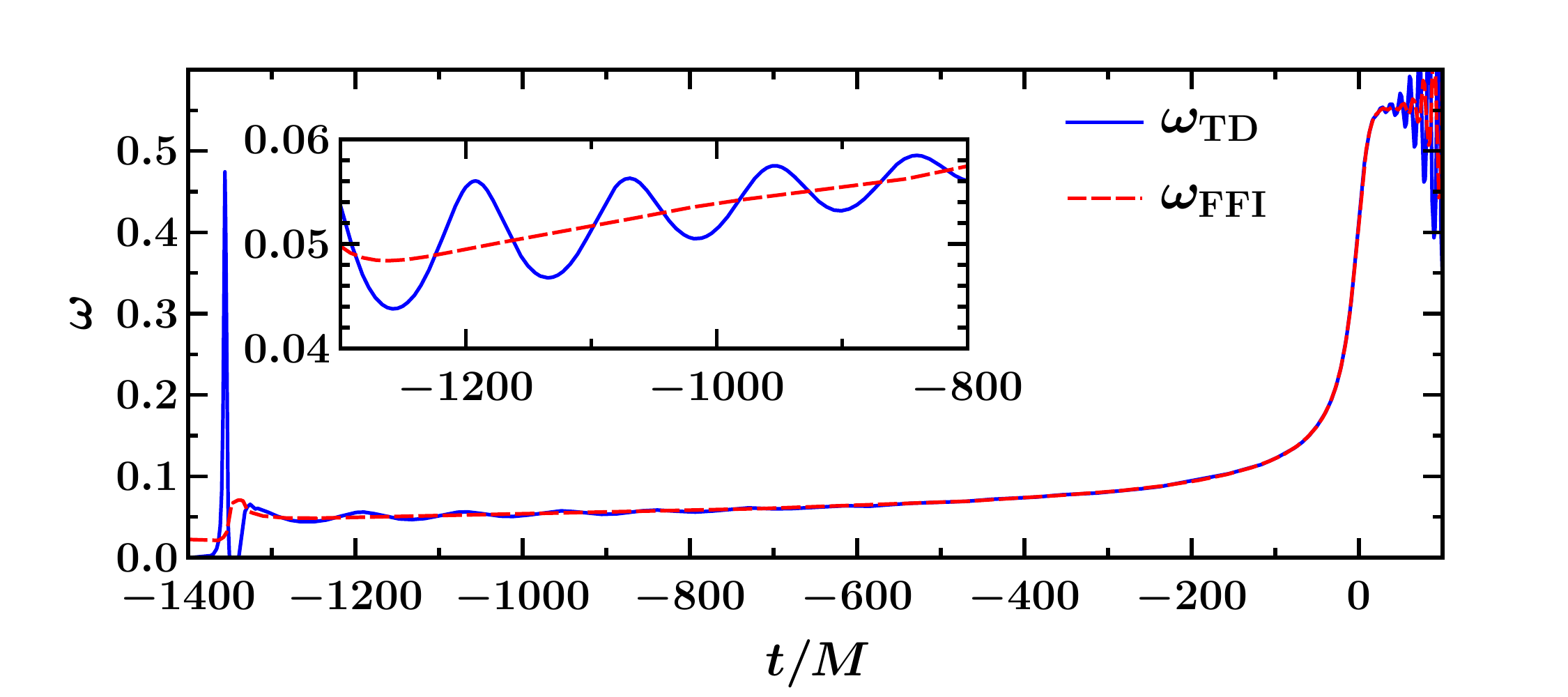}
  \caption{
    The instantaneous frequency $\omega$ derived from the time-domain
    integrated $(\ell,m)=(2,2)$ wave-mode (blue, solid line) and from
    the same mode integrated using the FFI scheme (red, dashed
    line). The spurious oscillations are essentially removed from
    the FFI curve, as is clear from the inset. (Note, however, the
    quality of the ringdown frequency remains poor as compared to the
    original $\psi_4$ data~\cite{Pollney:2009yz}.)
  }
  \label{fig:inst-freq}
\end{figure}

Reduced artificial oscillations are also apparent in other physically
important quantities. For instance, Fig.~\ref{fig:inst-freq} plots the
instantaneous frequency, $\omega(t)$ of the integrated $h_{22}$
computed in the time domain and via FFI. Artificial oscillations in
this quantity can be confused with physical eccentricity. Indeed, the
FFI result retains small oscillations in $\omega$, which are
consistent with those seen in the raw $\psi_4$ data, suggesting that
the small remaining physical eccentricity modes have been retained,
while the artificial non-linear drifts are removed.

Finally, we note that while the artificial integration drifts have a
notable visible effect, they seem to have little influence on
the use of $h$ in a detector template. We can compute
the \emph{match}
\begin{equation}
  \mathcal{M}[h_1,h_2]
  = \max_{t_0}\max_{\phi_1}\max_{\phi_2}
  \frac{\langle h_1|h_2\rangle}{\sqrt{\langle h_1|h_1\rangle \langle 
  h_2|h_2\rangle}}\,,
\end{equation}
between the time-domain integrated $h_{\rm TD}$ and $h_{\rm FFI}$, 
where
\begin{equation}
  \langle h_1|h_2\rangle = \int_{-\infty}^{\infty}df \frac{\th_1^*(f)
    \th_2(f)}{S_f(f)}\,,
\end{equation}
is a detector dependent scalar product involving the sensitivity curve
$S_f(f)$.

We find that the mismatch, $\mathcal{M}_{\rm mis} =
\mathcal{M} - 1$, between time-domain integration and FFI is never greater
than $5\times 10^{-4}$ for systems less than $500M_\odot$ for Advanced-LIGO,
and $5\times 10^{8}M_\odot$ for LISA. Thus, the effect of integration
drift over the length scales considered here is negligible in terms of
detection. Nevertheless, it is crucial for constructing appropriate
long templates by matching to post-Newtonian models \cite{Hannam:2010ky, MacDonald:2011ne}.

\section{Conclusion}

Transforming the variables commonly output by a numerical
simulation to the gravitational-wave strain
involves some fundamental uncertainties. These artifacts are a result
of the integration of finite length, discretely sampled, noisy data.
Independent sources of error contribute to large secular non-linear drifts in the
integrated data, in particular random-walk effects for time-domain
integrations, and spectral leakage in the frequency domain. These
effects have nothing to do with the simulation itself, i.e., they are
unrelated to gauge or local measurement effects. They are inevitable
regardless of the quality of the model (though lowering the level of
noise will reduce the effects of random-walk). And they are independent
of the genuine integration constants which also arise, but lead at
most to a linear drift. The simple prescription which
we have developed, FFI, which, given a single
parameter $\omega_0$, can be applied to any (oscillatory) $(\ell,m)$ waveform mode
(see \cite{Reisswig:2010cd} for an application in stellar core collapse)
suppresses the worst
of the problems in the analytic test-cases and numerous practical
examples which we have studied, including various spin configurations
of binary black hole inspiral. The method involves a
single parameter choice, $\omega_0$, and removes the bulk of the effect of spectral
leakage while maintaining the amplitude of all oscillation
modes. A similar effect can be achieved through a careful choice of
band-pass filters, though our experience suggests a certain amount of
experimentation is required before a similarly satisfactory result can
be obtained. This may be impractical for use in parameter space studies
involving a large number of physical models and modes.

The issue remains that removing the spurious non-linear drifts associated
with integration involves a modification of the data, and in
particular potential distortions of low-frequency physical
information. Especially, low-frequency information such as the linear and 
non-linear memory effect will be very hard to disentangle from the
``artificial memory'' induced by numerical error, since both effects
appear as low-frequency drifts in the time domain waveforms.
Also, without an exact solution to compare against, it is not
possible to arrive at a rigorous estimate of the error in the strain
calculated for a given waveform. The examples here suggest that a
variation on the order of $1\%$ in amplitude can be expected between
different integration methods or parameter choices. This source of
error should be taken into account, for instance, in matching
post-Newtonian results to numerically calculated strains for the
merger. 

Finally, we note that while we have emphasized that low-frequency
artifacts arise purely due to the process of numerical integration,
systematic aspects of the data measurement can still complicate the
situation. The results we have presented here use characteristic data
measured at $\scri$, and are free of the coordinate effects discussed
in the introduction.  Finite-radius extraction can be problematic in a
number of ways, related to the local gauge and dynamics of the
measurement sphere.  But also, specific truncation errors and numerical
artifacts may be
poorly correlated between measurements at different radii,
complicating the extrapolation of integrated quantities. We generally find
it preferable to extrapolate $\psi_4$ and then integrate. The secular
drifts in $h$ from extrapolated waveforms tend to be more problematic
than the characteristic results presented here.  However,
high-pass filter techniques and FFI are quite effective for such data
as well, though with an increased (and difficult to estimate)
systematic error due to the finite-radius effects.

\ack

The authors would like to thank Sascha Husa, Christian D.~Ott and
Ulrich Sperhake for helpful input. This work is supported by the
Bundesministerium f\"ur Bildung und Forschung and the National Science
Foundation under grant numbers AST-0855535 and OCI-0905046. DP has
been supported by grants CSD2007-00042 and FPA-2007-60220 of the
Spanish Ministry of Science.  Computations were performed on the NSF
Teragrid (allocation TG-MCA02N014), the LONI network
(\texttt{www.loni.org}) under allocation \texttt{loni\_numrel05}, at LRZ M\"unchen, the Barcelona
Supercomputing Center, and at the Albert-Einstein-Institut.


\section*{References}
\bibliographystyle{unsrt}
\bibliography{references}

\end{document}